\documentclass[aps,prl,twocolumn,amsmath,groupedaddress,showpacs]{revtex4}

\usepackage{graphicx}

\begin{document}

\title{Mode-Locking and Mode-Competition in a Non-equilibrium
Solid-State Condensate}

\author{P. R. Eastham}

\affiliation{Blackett Laboratory, Imperial College London, SW7 2BW, United Kingdom}

\date{\today}

\begin{abstract}
A trapped polariton condensate with continuous pumping and decay is analyzed
using a generalized Gross-Pitaevskii model. Whereas an equilibrium condensate
is characterized by a macroscopic occupation of a ground state, here the
steady-states take more general forms. Some are characterized by a large
population in an excited state, and others by large populations in several
states. In the latter case, the highly-populated states synchronize to a
common frequency above a critical density. Estimates for the critical density
of this synchronization transition are consistent with experiments.
  \end{abstract}

\pacs{71.36.+c, 71.35.Lk, 03.75.Kk, 42.55.Sa}

\maketitle

Recent experiments\ \cite{KRK+06,DPG+06,BGS+07} have provided
substantial evidence for a new type of Bose-Einstein condensate,
formed from polaritons in semiconductor microcavities\
\cite{SFW98}. Although in many respects these results parallel those
of condensation in atomic gases, the similarities conceal some
fundamental differences. In particular, the lifetime of a polariton is
typically only a few picoseconds, and is less than the lifetime of the
condensate\ \cite{DPG+06}. The condensate is therefore a
non-equilibrium steady-state, in which the decaying polaritons are
continually being replenished. 

Several consequences of this non-equilibrium aspect of the system have
now been predicted, based on both microscopic calculations and
generalizations of the Gross-Pitaevskii approach\
\cite{SKL06,WC07,jmjkcgle,lagoudakisvortices}. While
these predictions are undoubtedly interesting, more dramatic
departures from the physics of equilibrium condensates are seen
experimentally. An equilibrium condensate is characterized by a
macroscopic occupation at the chemical potential; for a trapped, ideal
Bose gas this is the ground-state energy of the trap, while more
generally it is the lowest eigenvalue of the Gross-Pitaevskii
equation. In some experiments there is indeed evidence for
equilibrated polariton distributions and large occupations of the
$\vec{k}=0$ ground-state of untrapped polaritons\
\cite{KRK+06,DPG+06,BGS+07}, but in others there are massive
occupations associated with excited states. This has been observed in
pillar traps\ \cite{bajoni-2007}, lattices\ \cite{lai2007}, and
disorder\ \cite{baas-2007,krizhan2}. Such results have been
argued to be evidence of polariton lasing\ \cite{bajoni-2007,IRP+96},
distinct from condensation, but this distinction is clear only to the
extent that condensation is restricted to equilibrium.

The aim of this paper is to outline a theory of the trapped polariton
condensate with pumping and decay. The treatment is based on a
recently-proposed generalized Gross-Pitaevskii equation (gGPE)\
\cite{jmjkcgle,WC07}. Motivated by the recent experiments, we go
beyond the assumption\
\cite{jmjkcgle,WC07,SKL06,lagoudakisvortices,shelykh:066402,malpuech:206402}
of condensation at a single ground-state energy, and systematically
investigate how the dynamics of the pumped system can lead to more
general forms. We first consider a small trap at low densities, and
show that the gGPE reduces to kinetic equations for the occupations of
the trap states. The occupations of the states are determined by gain
and loss processes which are independent of energy, {\it i.e.} a
laser-like mode competition. Thus the steady-state solutions can
include massive occupations of an excited state, or of several states
simultaneously; the character of the steady-states in this limit can
be predicted from the wavefunctions of the trap.

Having established the steady-state structure at low densities, we then
numerically explore how it changes away from this limit. We demonstrate that,
if the trap is such that there are several massively-occupied states at low
densities, then above a threshold density the steady-state reverts to a
massive occupation at a single energy. This can be understood as the classical
phenomenon of synchronization (mode-locking) of coupled nonlinear oscillators\
\cite{pikovsky01}. Supposing that the dominant nonlinearity is the
polariton-polariton interaction we predict that, for realistic parameters,
polariton systems could be tuned through the synchronization transition. They
thus give access to both an interaction-dominated regime, and a regime of
laser-like mode competition.

Synchronization in the polariton condensate has been independently
considered by Wouters\ \cite{wouterspc}. That treatment is
complementary to this one, considering the specific problem of a
double-well trap using a related model. The double-well has also been
considered under resonant pumping\ \cite{sarchi07pp}, which further
differs from the present work because the condensate is directly
induced by the pump.

We consider the order-parameter equation\
\cite{jmjkcgle}\begin{equation} i\partial_t \psi = \left[H_0 + U
    |\psi|^2+i (\gamma_{\mathrm{eff}}(r)-\Gamma |\psi|^2)\right]\psi.
\label{eq:cgle}\end{equation} This complex
Ginzburg-Landau form\ \cite{cglermp} has been argued\
\cite{jmjkcgle,WC07} to be the generic mean-field description of a
polariton condensate with continuous pumping and decay. The first two
terms on the right form the standard Gross-Pitaevskii mean-field
equation, with local density $|\psi|^2$. $H_0$ includes a quadratic
approximation to the polariton dispersion, and a potential due to
disorder or the trap. $U>0$ is the strength of the repulsive
interaction between polaritons, which is treated as a contact
interaction because its range is of the order of the exciton Bohr
radius. The polarization degeneracy of the polaritons has been
neglected for simplicity; it could be treated using a two-component
order-parameter equation\ \cite{shelykh:066402}.

The last two terms on the right of (\ref{eq:cgle}) account for the
pumping and decay. The pumping model involves a reservoir of
high-energy particles, created by some external excitation. The
condensate is populated by stimulated scattering from this reservoir,
contributing a linear gain term $i\gamma(r)\psi$, where $\gamma(r)$ is
related to the reservoir density. This term combines with a similar
term from the decay of the polaritons, giving the overall linear gain
term, with coefficient $\gamma_{\mathrm{eff}}$, in
(\ref{eq:cgle}). However, the rate of condensate growth should reduce
with increasing density, as the pump reservoir becomes depleted.  This
effect, modeled using a single gain-saturation coefficient $\Gamma$,
gives the final term in (\ref{eq:cgle}). Physically, this form of
pumping can be interpreted in terms of a pump which tries to locally
enforce a steady-state density $\gamma_{\mathrm{eff}}(r)/\Gamma$.

\emph{Strong-Trapping Limit--} The steady-states of (\ref{eq:cgle})
can be determined analytically in the limit of strong trapping, where
the nonlinearities are weak compared with the single-particle level
spacing. We may then treat them with degenerate perturbation
theory. We expand $\psi(r,t)$ in terms of the eigenstates of $H_0$,
$\psi^0_n$, and retain only resonant terms in the resulting
equations. To simplify the notation we analyze a trap with only two
single-particle states, so
$\psi(r,t)=\mu_1(t)\psi^0_1(r)+\mu_2(t)\psi^0_2(r)$, and assume
homogeneous pumping. The amplitude $\mu_{1}(t)$ ($\mu_{2}(t)$) obeys
\begin{eqnarray} i\dot\mu_{1(2)}&=&\left[E_{1(2)}+i\gamma_{\mathrm{eff}}+(U-i\Gamma)\right. \nonumber \\ &&\left. \times(\eta_{1(2)} |\mu_{1(2)}|^2+2b |\mu_{2(1)}|^2)\right]\mu_{1(2)}. \label{eq:twomoderes} \end{eqnarray}
$E_i$ is the single-particle energy, and the wavefunctions have been
taken to be normalized and real.  $\eta_1=u_{1111}, \eta_2=u_{2222},$
and $b=u_{1122}$ are matrix elements for a local nonlinearity,
\begin{equation} u_{ppqq}=\int (\psi^0_p)^2 (\psi^0_q)^2
dr. \label{eq:matrixelements}\end{equation} They parametrize the
inhomogeneous density profile of the trap states, with $\eta_{1}$ and
$\eta_{2}$ describing the inhomogeneity of the states, and $b$ their
overlap.

Introducing number and phase variables in a rotating frame,
\begin{equation}\mu_{1(2)}(t)=e^{-i\gamma_{\mathrm{eff}}
    U/\Gamma}\sqrt{n_{1(2)}(t)}e^{-i\phi_{1(2)}(t)}, \end{equation}
separates the number and phase dynamics. The former obeys the rate
equations
\begin{equation}\label{eq:denseqs} \dot n_{1(2)}=2\Gamma
n_{1(2)}(\gamma_{\mathrm{eff}}/\Gamma-\eta_{1(2)}n_{1(2)}-2bn_{2(1)}),
\end{equation} with terms describing the stimulated scattering from
the reservoir and the spontaneous decay, and the reservoir
depletion. This result can be understood as a generalization of the
kinetic description of polariton lasing\
\cite{PhysRevB.66.085304,PhysRevB.59.10830,PhysRevB.65.153310,sarchi:115326},
to treat the spatial structure of the trap. It describes the extensive component
of the occupation, and hence effects such as spontaneous pumping are
missing. They would be important for a finite system close to
threshold\ \cite{eastham:085306}. 

The phase dynamics is straightforward, obeying \begin{equation}
\dot\phi_{1(2)}=E_{1(2)}-U(\gamma_{\mathrm{eff}}/\Gamma-\eta_{1(2)}n_{1(2)}-2bn_{2(1)}). \label{eq:phasedynamics}\end{equation}
Each mode oscillates freely, at a single-particle energy which is shifted by
the repulsive interactions.

The rate equations (\ref{eq:denseqs}) have several steady-state solutions above the bulk
condensation threshold, $\gamma_{\mathrm{eff}}=0$. There are always
two solutions corresponding to condensation in each of the trap
states: either $n^s_1=\gamma_{\mathrm{eff}}/(\eta_1\Gamma)$ is finite
and $n^s_2=0$ vanishes (state $S_1$), or \emph{vice versa} (state
$S_2$).  However, if either $\eta_{1},\eta_{2}>2b$ or
$\eta_{1},\eta_2<2b$ there is also a steady-state $T$, with massive
occupations of both trap states:
 \begin{equation}
 n_{1(2)}=\frac{\gamma_{\mathrm{eff}}}{\Gamma}\left(\frac{2b-\eta_{2(1)}}{4b^2-\eta_1\eta_2}\right).
\label{eq:twomodeamplitudes}\end{equation} 

\begin{figure}[t]
\includegraphics{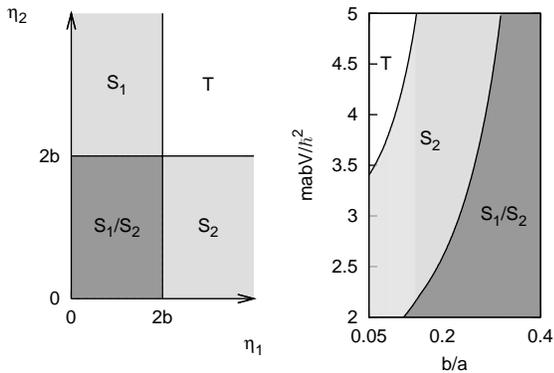}
\caption{Left panel: steady-state structure for condensation in a two-state
  trap, for the generalized Gross-Pitaevskii model neglecting non-resonant
  terms (\ref{eq:twomoderes}). Lettering denotes the stable steady-states in
  each region, with condensation in both trap states (T), state 1 alone
  ($S_1$), or state 2 alone ($S_2$). In the lower-left region both $S_1$ and
  $S_2$ are stable, and the steady-state is selected by the initial
  conditions. Right panel: steady-state structure for the two lowest states of
  a one-dimensional model trap, consisting of a hard-wall trap of length
  $a$ with a trap of length $b$ and depth $V$ at its
  center. \label{fig:phasediagram} }\end{figure}

The left panel of Fig.\ \ref{fig:phasediagram} shows the parameter regions in
which the different steady-states are stable. Linearizing (\ref{eq:denseqs})
about the steady-state $S_1$, we find that population fluctuations decay with
rates $\lambda_1=2\gamma$ and $\lambda_2=2\gamma(2b/\eta_{1}-1)$. Thus this
solution is stable for $\eta_1<2b$. This is the condition that the occupation
of the first trap state alone, determined by the pumping and the
self-gain-saturation parameter $\eta_1$, is sufficient to keep the second
below threshold. The analogous condition $\eta_2<2b$ holds for the existence
of a stable solution in which only the second trap state is occupied,
$S_2$. If neither criterion is satisfied both states must be occupied, and
this is where the two-mode solution $T$ is stable -- fluctuations there decay
with rates $2\gamma$ and $2\gamma(2b-\eta_1)(2b-\eta_2)/(4b^2-\eta_1\eta_2)$.

The different steady-state solutions can be realized in many different
potentials. For two widely-separated states we generically have $b\to 0$, and
hence obtain two coexisting condensates. While this result is perhaps expected
for separated traps, it can nonetheless also occur when there is substantial
spatial overlap. A simple model which demonstrates this is a one-dimensional
hard-wall trap of size $a$, with a finite trap of size $b$ and depth $V$ at
its center. The right panel of Fig.\ \ref{fig:phasediagram} shows the
different steady-state regions for the two lowest eigenstates of this
potential.

Experimentally, the presence of massively-occupied states in polariton
systems is shown by bright luminescence peaks. The characteristic
frequencies in such optical spectra follow from
(\ref{eq:phasedynamics}). In $S_{1}(S_2)$ the massively-occupied
states will lead to strong emission at the frequency
$U\gamma_{\mathrm{eff}}/\Gamma+E_{1(2)}$, while in $T$ both peaks
appear simultaneously. Note that here the blueshift of the condensing
modes is $U\gamma_{\mathrm{eff}}/\Gamma$, irrespective of which
steady-state we consider. Physically, this is because the energy
shifts are determined by the density, which is fixed by the
pumping. 

In addition to the strong emission associated with the condensing
states, we also expect peaks in the optical response associated with
the non-condensing trap states. For the mean-field model in the
resonant approximation these states are not populated, and hence would
appear in the absorption but not the luminescence. However, they could
develop non-macroscopic populations due to effects beyond that model,
in which case they would appear weakly in luminescence. These peaks
are shifted by the mean-field repulsion with the condensate, so in
$S_{1}$ the absorption peak will be at $E_{2}+2bn_{1}$. Since its
width will be proportional to $\lambda_2$, the mode is narrow close to
the critical line $\eta_{1}=2b$, where it goes unstable.

\begin{figure}[t]
\includegraphics{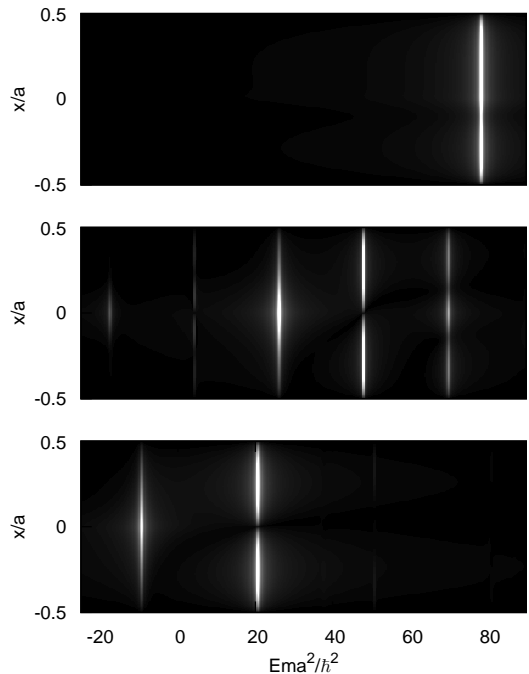}
\caption{Spectral analysis of the polariton field in the steady-states of the
  generalized Gross-Pitaevskii model, with the potential described in the
  caption of Fig.\ \ref{fig:phasediagram}. The pump strength increases through
  $\gamma_{\mathrm{eff}}=1, 30, 60$ from the lowest panel to the highest.  The
  grayscale is the computed amplitude $|\psi(E,x)|$, normalized by
  $\sqrt{\gamma_{\mathrm{eff}}}$ to account for the overall increase in
  density. $U/\Gamma=1$, $b=0.05$ and $V=25$.\label{fig:pumpnr}}\end{figure}

\emph{Beyond Strong-Trapping--} Having established the physics of
(\ref{eq:cgle}) in the strong-trapping regime, we now investigate how it
develops with increasing nonlinearity. To do this we use real-space
discretization to solve (\ref{eq:cgle}) directly, for the one-dimensional
model described above. This potential is chosen as a simple,
few-parameter model, which can be tuned into the different steady-state
regions. We choose $a=m=\Gamma=1$, taking $a$ as the unit of length,
$\hbar^2/(ma^2)$ as the unit of energy, and working with a rescaled density
$|\psi|^2\to|\psi|^2/\Gamma$. Both the repulsive interaction $U$ and the gain
saturation $\Gamma$ are believed to be significant for polaritons\
\cite{KRK+06,jmjkcgle,WC07}, and so for this demonstration we take
$U/\Gamma=1$.

Fig.\ \ref{fig:pumpnr} illustrates how the steady-state spectra
develop with increasing pumping. The analysis above, applied to the
two lowest states of this potential, predicts the two-mode
steady-state $T$. This is in agreement with the results at weak
pumping (lowest panel). Increasing the pumping there is the overall
blueshift associated with the increased density. The lowest two modes
still dominate the spectrum, but their energy splitting has reduced
slightly, and several further emission peaks appear (middle
panel). Further increasing the pumping, the spectrum switches to
emission at a single frequency (top panel).

At a general level, these results are expected consequences of the
non-resonant terms dropped from (\ref{eq:twomoderes}). In a two-mode model
with states of opposite parity, for example, there is the additional Josephson
term\begin{equation}b(U-i\Gamma)\mu_{1(2)}^\ast\mu_{2(1)}\mu_{2(1)}\label{eq:nlcouple}\end{equation}
in the equation (\ref{eq:twomoderes}) for $\mu_{1}$($\mu_{2}$). Thus the
condensing states can drive nonlinear emission at other frequencies, as we see
at intermediate pumping. Furthermore, the non-resonant terms couple together
the phases of the condensing modes in (\ref{eq:phasedynamics}). Thus we expect
frequency pulling, and eventually synchronization of the condensing states\
\cite{pikovsky01}.

Coupled oscillators synchronize to a common frequency when the
phase-phase couplings become comparable to their energy splitting
$\Delta$. Since here the states have similar and overlapping density
profiles, the matrix elements (\ref{eq:matrixelements}) are all of
order one. The scale of the phase-dependent couplings in
(\ref{eq:phasedynamics}) is then set only by the nonlinearity
$\sqrt{U^2+\Gamma^2}$ and the polariton density $\rho\approx
\gamma_{\mathrm{eff}}/\Gamma$. We may thus estimate the critical
polariton density for the mode-locking transition, $\rho_c$, from
\begin{equation}\Delta\sim \rho_c \sqrt{U^2+\Gamma^2}\sim
\gamma_{\mathrm{crit}}\sqrt{1+U^2/\Gamma^2}.\end{equation} This form
is consistent with Fig.\ \ref{fig:pumpnr}, and with simulation results
(not shown) for other values of the nonlinearities $U$ and
$\Gamma$. In particular, in the same model the synchronization occurs
between $\gamma_{\mathrm{eff}}/\Gamma=20$ and $40$ when $U=0$, and
between $U/\Gamma=20$ and $40$ at fixed
$\gamma_{\mathrm{eff}}/\Gamma=1$.

Synchronization due to nonlinear gain is well-known as the basis for
mode-locked lasers. However, an important difference between the polariton
condensate and a laser is the presence of strong interactions $U$ between the
particles, due to the excitonic component of the polariton. Thus
frequency-pulling and synchronization could be expected to have a much wider
role in the physics of polariton condensates than they do in lasers, occurring
on large energy scales at low intensities.

Let us estimate the critical polariton density $\rho_c$ for
synchronization, supposing it is controlled by the real nonlinearity
$U$. Estimates of this interaction are available for plane-wave
excitons in a perfect quantum-well, and the localized exciton states
of a disordered quantum well\ \cite{KMS+07}. For current experiments,
a plausible upper limit is the result $U\sim \Omega_{R}(m_x W)^{-1}$
from the disordered models, with disorder energy-scale $W\sim
1\mathrm{meV}$. $m_x \sim 0.5 m_e$ is the exciton mass, and
$\Omega_{R}\sim 20 \mathrm{meV}$ the Rabi splitting. Thus the phase
boundary for synchronization in a trap of scale $L_t$ is
\begin{equation} \Omega_{R}(m_x W)^{-1} \rho_c \sim 1/(m
L_t^2). \label{eq:phaselocking}\end{equation} For a trap of $L_t\sim 5
\mu \mathrm{m}$\ \cite{bajoni-2007} and a polariton mass $m\sim
10^{-5}m_e$ this gives $\rho_c \sim 10^{11} \mathrm{cm^{-2}}$, or
$n_c\sim L_t^2\rho_c \sim10^3$. This estimate is about one order of
magnitude larger than the densities usually reported for polariton
condensates. Thus, with a suitable potential, coexisting polariton
condensates of different frequencies should be expected in tight
traps. The synchronization transition could then be observed with
increasing density, if necessary using larger traps to reduce $\rho_c$
into the experimentally accessible range.  

Since we do not expect synchronization at current densities in tight
traps, the present theory is consistent with the observations there of
well-resolved emission lines\ \cite{bajoni-2007,krizhan2}. It
also appears to be consistent with the existence of long-range
coherence\ \cite{KRK+06,baas-2007} at current densities, with softer
traps formed by disorder.

\emph{Concluding Remarks--} Because the tightly-trapped polariton condensate
has well-resolved emission lines, it could provide a sensitive probe of the
physics of non-equilibrium condensates. In particular, the present results
will allow the order-parameter equation (\ref{eq:cgle}) to be tested. While
this form certainly captures much of the physics\ \cite{WC07,jmjkcgle},
effects which are important elsewhere are missing. Most obviously, there is no
thermalization with the reservoirs\ \cite{PhysRevLett.82.3927,CDT+04}, which
exchange particles with the system irrespective of energy, and do not directly
cause transitions between trap states. Such effects would appear in
generalizations of the kinetic equations (\ref{eq:denseqs}). Since some
groups\ \cite{KRK+06,DPG+06} report thermalized distributions, such
generalizations may prove necessary.

To conclude, we have considered the trapped polariton condensate in
the framework of a generalized Gross-Pitaevskii model. At low
densities, this model admits solutions which, differently from an
equilibrium condensate, involve massive occupations of excited states,
or of several states simultaneously. We have derived criteria for
predicting the nature of the steady-states in a given geometry, and
shown that the steady-states are selected by gain-competition
effects. At a general level such physics is of course familiar in
lasers, though it has not previously been considered for the polariton
system. Moreover, the direct interactions between polaritons create
differences compared with the photon laser: blueshifting the modes in
the weak-nonlinearity regime, and causing frequency-pulling and
synchronization at stronger nonlinearities. Our estimates for the
critical density of the synchronization transition suggest that it
could be cleanly observed in tight traps\ \cite{bajoni-2007}, and may
be responsible for the observations of long-range coherence across
disorder potentials\ \cite{KRK+06,baas-2007}.

This work was supported by EPSRC EP/C546814/01. I acknowledge helpful
correspondence with M. Wouters.


\end{document}